\documentclass[twocolumn]{aastex62}

\usepackage{hyperref}

\newcommand{\ltsima}{$\; \buildrel < \over \sim \;$}
\newcommand{\simlt}{\lower.5ex\hbox{\ltsima}}

\def\arcdeg{\hbox{$^\circ$}}
\def\arcmin{\hbox{$^\prime$}}
\def\arcsec{\hbox{$^{\prime\prime}$}}

\newcommand{\aref}[1]{\hyperref[#1]{Appendix~\ref{#1}}}

\shorttitle{WR~20a X-Ray Light Curve}
\shortauthors{Olivier et al.}

\begin{document}

\title{A Multiwavelength Study of the Massive Colliding Wind Binary WR~20a: A Possible Progenitor for Fast-Spinning LIGO Binary Black Hole Mergers}

\author[0000-0002-4606-4240]{Grace M. Olivier}
\affil{Department of Physics and Astronomy, Texas A\&M University, 4242 TAMU, College Station, Texas 77843, USA}
\affil{George P. and Cynthia Woods Mitchell Institute for Fundamental Physics and Astronomy, Texas A\&M University, 4242 TAMU, College Station, Texas 77843, USA}
\affil{Department of Astronomy, The Ohio State University, 140 W. 18th Ave., Columbus, Ohio 43210, USA}
\affil{Center for Cosmology and AstroParticle Physics, The Ohio State University, 191 W. Woodruff Ave., Columbus, OH 43210, USA}

\author[0000-0002-1790-3148]{Laura A. Lopez}
\affil{Department of Astronomy, The Ohio State University, 140 W. 18th Ave., Columbus, Ohio 43210, USA}
\affil{Center for Cosmology and AstroParticle Physics, The Ohio State University, 191 W. Woodruff Ave., Columbus, OH 43210, USA}
\affil{Flatiron Institute, Center for Computational Astrophysics, NY 10010, USA}

\author[0000-0002-4449-9152]{Katie Auchettl}
\affil{OzGrav, School of Physics, The University of Melbourne, Parkville, Victoria 3010, Australia}
\affil{ARC Centre of Excellence for All Sky Astrophysics in 3 Dimensions (ASTRO 3D)}
\affil{Department of Astronomy and Astrophysics, University of California, Santa Cruz, CA 95064, USA}

\author[0000-0003-4423-0660]{Anna L. Rosen}
\affil{Center for Astronomy \& Space Sciences, University of California San Diego, La Jolla, CA 92093, USA}

\author[0000-0002-3269-3847]{Aldo Batta}
\affil{Instituto Nacional de Astrofísica, Óptica y Electrónica, Tonantzintla, Puebla 72840, México}
\affil{Consejo Nacional de Ciencia y Tecnolog\'ia, Av. Insurgentes Sur 1582, 03940, Ciudad de M\'exico, M\'exico}

\author[0000-0002-5787-138X]{Kathryn F. Neugent}
\altaffiliation{NASA Hubble Fellow}
\affil{Dunlap Institute for Astronomy \& Astrophysics, University of Toronto, Toronto, ON M5S 3H4, Canada}
\affil{Center for Astrophysics, Harvard \& Smithsonian, 60 Garden St., Cambridge, MA 02138, USA}

\author[0000-0003-2558-3102]{Enrico Ramirez-Ruiz}
\affil{Department of Astronomy and Astrophysics, University of California, Santa Cruz, CA 95064, USA}
\affil{DARK, Niels Bohr Institute, University of Copenhagen, Denmark}

\author[0000-0002-6244-477X]{Tharindu Jayasinghe}
\altaffiliation{NASA Hubble Fellow}
\affil{Department of Astronomy,  University of California Berkeley, Berkeley CA 94720, USA}
\affil{Department of Astronomy, The Ohio State University, 140 W. 18th Ave., Columbus, Ohio 43210, USA}
\affil{Center for Cosmology and AstroParticle Physics, The Ohio State University, 191 W. Woodruff Ave., Columbus, OH 43210, USA}

\author[0000-0001-5661-7155]{Patrick J. Vallely}
\affil{Department of Astronomy, The Ohio State University, 140 W. 18th Ave., Columbus, Ohio 43210, USA}
\affil{Center for Cosmology and AstroParticle Physics, The Ohio State University, 191 W. Woodruff Ave., Columbus, OH 43210, USA}

\author[0000-0003-2431-981X]{Dominick M. Rowan}
\affil{Department of Astronomy, The Ohio State University, 140 W. 18th Ave., Columbus, Ohio 43210, USA}
\affil{Center for Cosmology and AstroParticle Physics, The Ohio State University, 191 W. Woodruff Ave., Columbus, OH 43210, USA}

\email{gmolivier@tamu.edu}

\begin{abstract}

WR~20a is the most massive close-in binary known in our Galaxy. It is composed of two $\approx$80~M$_\odot$ Wolf-Rayet stars with a short period of $\approx$3.7 days in the open cluster Westerlund 2. As such, WR~20a presents us with a unique laboratory for studying the currently uncertain physics of binary evolution and compact object formation as well as for studying the wind collision region in an massive eclipsing binary system. We use deep \textit{Chandra} observations of WR~20a to study the time variability of the wind collision region between the two Wolf-Rayet stars and are able to produce an X-ray light curve covering $\approx$2/3 of its orbital period. We find that the X-ray light curve is asymmetric because the flux of one peak is 2.5$\sigma$ larger than the flux of the other peak. This asymmetry could be caused by asymmetric mass-loss from the two stars or by the lopsidedness of the wind collision region due to the unusually fast rotation of the system. The X-ray light curve is also shifted in phase space when compared to the optical light curves measured by \textit{TESS} and ASAS-SN. Additionally, we explore the ultimate fate of this system by modeling the resultant  binary black hole merger expected at the end of the two stars' lives. We conclude that this system will evolve to be a representative of the sub-population of LIGO progenitors of fast-spinning binary black hole merger events.

\end{abstract} 


\section{Introduction}

WR~20a is a massive binary system located $\approx $1 pc away from the center of Westerlund 2, one of the most massive star clusters ($\approx10^{4}~M_{\odot}$; \citealt{ascenso07}) in the Milky Way. Westerlund 2 is a young open cluster at only $\lesssim$2~Myr old, and contains $\approx$30 O-type stars \citep{rauw07,vargas13} and two Wolf-Rayet (WR) binary systems, WR~20a and WR~20b \citep{vanderhucht01}. WR~20a is an uncommon close-in binary system as it is composed of nearly identical WN6ha stars of mass $82+83~M_{\odot}$ in a circular orbit with a period of $\approx$3.7~days \citep{bonanos04,rauw05}. The WN6ha spectral type denotes that these WR stars have the enhanced nitrogen characteristic of a WN-type WR but additionally show hydrogen absorption lines in their spectra. Therefore, unlike most WR stars, these stars are not fully evolved objects and are instead still burning hydrogen on the main sequence and are $\approx$2 Myr old, consistent with the cluster age \citep{smith08}. As expected due to the high temperatures and masses, this system is very luminous when compared to other sources within the cluster, with $L_{\rm bol} \approx 2\times10^{39}$~erg~s$^{-1}$ and $L_{\rm X} \approx 1.4 \times10^{33}$~erg~s$^{-1}$ \citep{naze08}. 

The Wolf-Rayet stars in this binary are capable of producing extremely fast and dense winds \citep{maeder10,vink21}. These winds collide in the area between the stars in a region known as the wind collision region (WCR) where the material is shock heated to X-ray emitting temperatures due to their fast wind velocities \citep{parkin08,montes13,rosen2022}. Often times, these WCRs are observed in massive binaries with an O-star and a WR star in which case their X-ray light curves are asymmetric due to the higher mass-loss rate of the WR \citep[e.g.,][]{pollock18, garofali19, pollock21, pradhan21}. However, WR~20a is composed of near identical WR stars, so a symmetrical X-ray light curve is predicted from the WCR \citep{montes13} since these stars should have roughly identical wind properties \citep{rauw05, vink21}.

The ultimate fate of WR~20a is a tantalizing prospect to consider. These two massive stars are still on the main-sequence and appear to be chemically similar based on their spectra \citep{rauw05}. Since these stars have not turned off the main-sequence yet and are experiencing mass-loss stripping due to being in a close binary system, it is thought that these massive twins will likely evolve into Luminous Blue Variables (LBVs) after leaving the main sequence \citep{rauw05,smith08}. The mass loss involved in the LBV stage will likely dramatically change the mass of the star at the end of its life, moving it from candidates for pair-instablity supernovae \citep[PISN,][]{renzo20} 
to core collapse supernovae with black hole remnants. If these two massive stars leave behind black holes in an orbit similar to their current short-period orbit, their remnants may coalesce in a gravitational wave event similar to those detectable by LISA and the LIGO/VIRGO collaboration. The LIGO/VIRGO Collaboration has discovered numerous binary black hole (BBH) coalescences since its first discovery in 2015 \citep{abbott16a,abbott16b,ligo21a,ligo21b}. Due to its member masses, WR~20a could be an example of the massive binary pathway to BBH mergers. If WR~20a is a LIGO progenitor source, it might be a system that exhibits two long gamma-ray bursts (GRBs) associated with the supernova explosion from each star \citep{2004MNRAS.348.1215I}, and then coalesce in a LIGO detectable event \citep{bavera21}.

In this paper, we analyze multiwavelength data of the optical and X-ray light curves of WR~20a to characterize its WCR and interaction. We adopt a distance of $D = 4.2$~kpc to WR~20a based on the analysis of the Westerlund 2 stellar populations by \citet{vargas13}. We note that all luminosities from the literature quoted in this text have been recalculated assuming this distance. The outline of this paper is as follows. In \autoref{sec:data} we describe the analysis of data from \textit{Chandra}, ASAS-SN, and \textit{TESS}. In \autoref{sec:results} we present our results from the light curve analysis from the optical and X-ray data. In \autoref{sec:wcr} we discuss the implications for the WCR based on the X-ray light curve and in \autoref{sec:ligo} we discuss the feasibility of WR~20a tracing a pathway of LIGO sources. In \autoref{sec:conclusions} we present our conclusions.


\section{Data and Analysis} \label{sec:data}

In order to compare the X-ray variability to the orbital period, we use optical data to constrain the orbital period of WR~20a. We utilize data from the ground-based ASAS-SN survey and the \textit{TESS} satellite to understand the optical parameters of this binary which we will present in \autoref{sec:optdata}.

\subsection{X-rays and Spectral Analysis by Phase} \label{sec:xraydata}

We use data from the \textit{Chandra} X-ray Observatory to produce the X-ray light curve of the thermal photons that emanate from the WCR of WR~20a and are produced by the shock heating of the fast, colliding stellar winds.  We include both archival observations from 2006 (PI: Rauw; ObsID:6410,6411) of 98~ks and new observations taken in 2018 (PI: Lopez; ObsID:20133-34,21842,21847-48) of 268~ks.  The seven ACIS-I observations we include in this work are listed in \autoref{table:data}. The 2006 data are not the only previous observations of Westerlund 2. In the 2003 observation of RCW~49 (PI: Garmire; ObsID 3501) WR~20a falls on a chip gap and thus is not observed. Additionally, the 2012 observation of PSR J1022$-$5746 (PI: Garmire; ObsID 12151) does not resolve WR~20a as it is $\approx$7\arcmin\ off axis, so we do not include it in this work. 

\begin{figure}
\begin{center}
    \includegraphics[width=\columnwidth]{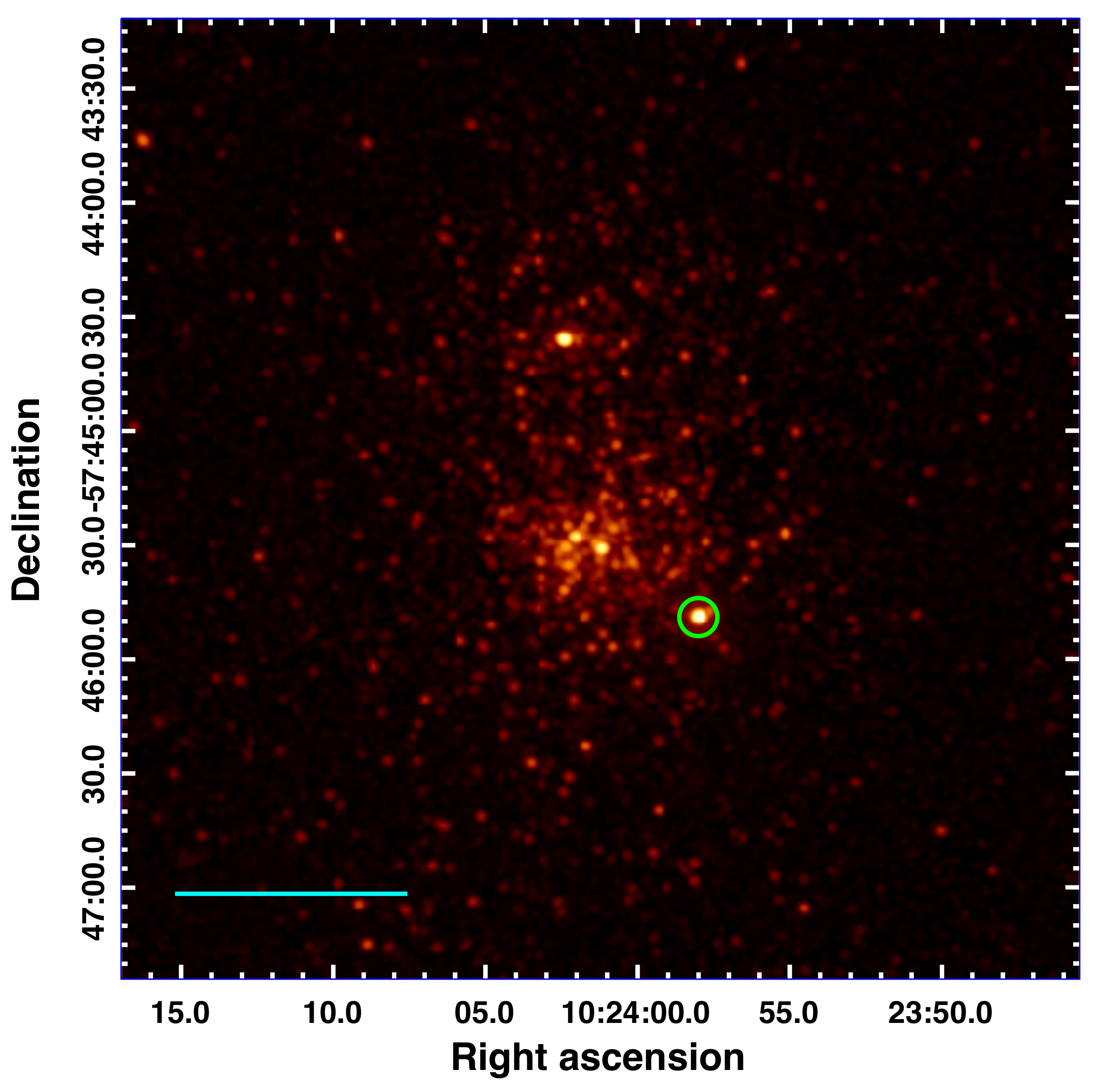}
\end{center}
\caption{Merged, exposure-corrected broad-band ($0.5-7.0$~keV) X-ray image of Westerlund 2 from 366~ks of \textit{Chandra} data. The green circle shows WR~20a, and the cyan scale bar denotes 1\arcmin. North is up, and East is left.}
\label{fig:xrayimg}
\end{figure}

We reduce the \textit{Chandra} data using \textit{Chandra} Interactive Analysis of Observations (\textsc{ciao}) Version 4.12 \citep{fru06}.  We employ the command \textit{merge\_obs} to produce the exposure-corrected broad-band ($0.5-7.0$~keV) image shown in \autoref{fig:xrayimg} of Westerlund 2. The green circle marks the location of WR~20a, and the cyan scale bar shows 1\arcmin. 


\begin{deluxetable}{lrcc}
\tablecolumns{4}
\tablewidth{0pt} \tablecaption{{\it Chandra} Observations \label{table:data}} 
\tablehead{\colhead{ObsID} & \colhead{Exposure} & \colhead{UT Start Date} & \colhead{$\log L_{\rm X}/L_{\rm bol}$}}  
\startdata
6410 & 49~ks & 2006-09-05 & -6.26\\
6411 & 49~ks & 2006-09-28 & -6.15\\
20133 & 26~ks & 2018-09-11 & -6.21\\
20134 & 33~ks & 2018-09-18 & -6.19\\
21842 & 57~ks & 2018-09-13 & -6.14\\
21847 & 73~ks & 2018-09-19 & -6.12\\
21848 & 79~ks & 2018-09-21 & -6.17
\enddata
\end{deluxetable}

\begin{figure*}
\begin{center}
    \includegraphics[width=\textwidth]{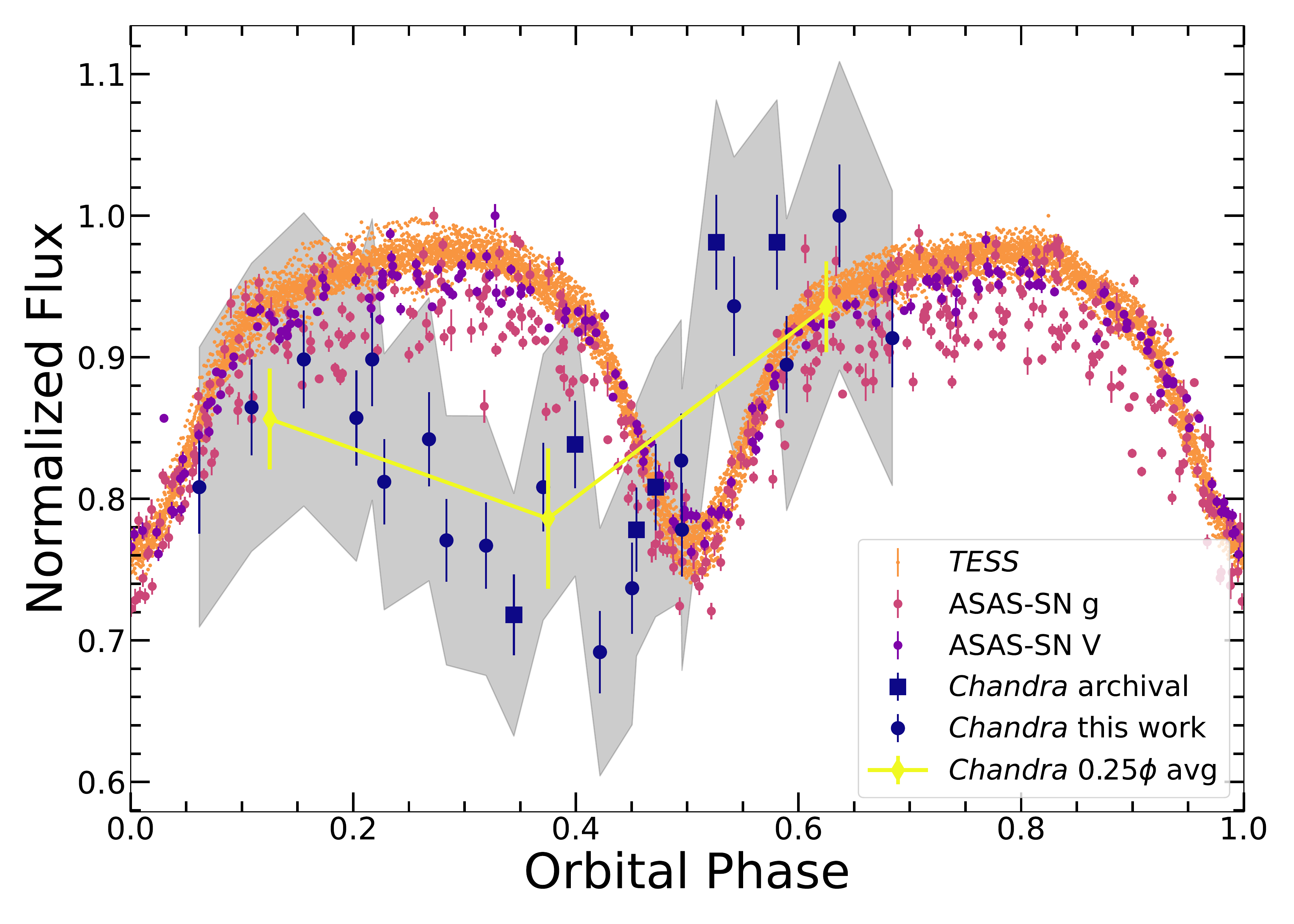}
\end{center}
\caption{Phase folded light curve for WR~20a. The \textit{TESS} data are shown in the orange small points, the \textsc{ASAS-SN} data from the g and V bands are shown in the pink and purple points respectively, and the \textit{Chandra} data are plotted as the navy circles for the new data and squares as the archival data with the 3$\sigma$ errors plotted as the gray band. The yellow diamonds show the average flux for bins of 0.25$\phi$ that are centered at $\phi$ = 0.125, 0.375, and 0.625 to capture the peaks and troughs of the X-ray light curve. The difference between the bins at the two peaks is 2.5$\sigma$.}
\label{fig:lightcurve}
\end{figure*}

Additionally, we use {\sc ciao} to extract spectra from the reprocessed event 2 files from both the new and archival observations. We create evenly spaced bins of $\approx$15~ks from each observation using \textit{dmcopy} from {\sc ciao} and extract spectra using the command \textit{specextract}. We required that these bins have $\gtrsim$500 net counts to ensure good spectral fits, and this requirement resulted in the $\approx$15~ks length of each bin. The 24 resulting bins range from 14.2~ks to 17.8~ks in duration, yielding $\approx$600$-$900 net counts, which are reported in \autoref{table:results}. When extracting spectra, we adopt a source region of radius 2.5\arcsec\ around WR~20a, and we select seven background regions that are 0.1-0.2\arcmin\ in radius which do not include any point sources in order to determine the background X-ray emission around Westerlund 2. The fluxes from these $\approx$15~ks bins are shown as the navy circles in \autoref{fig:lightcurve}.

\begin{figure*}
\begin{center}
    \includegraphics[width=0.32\textwidth]{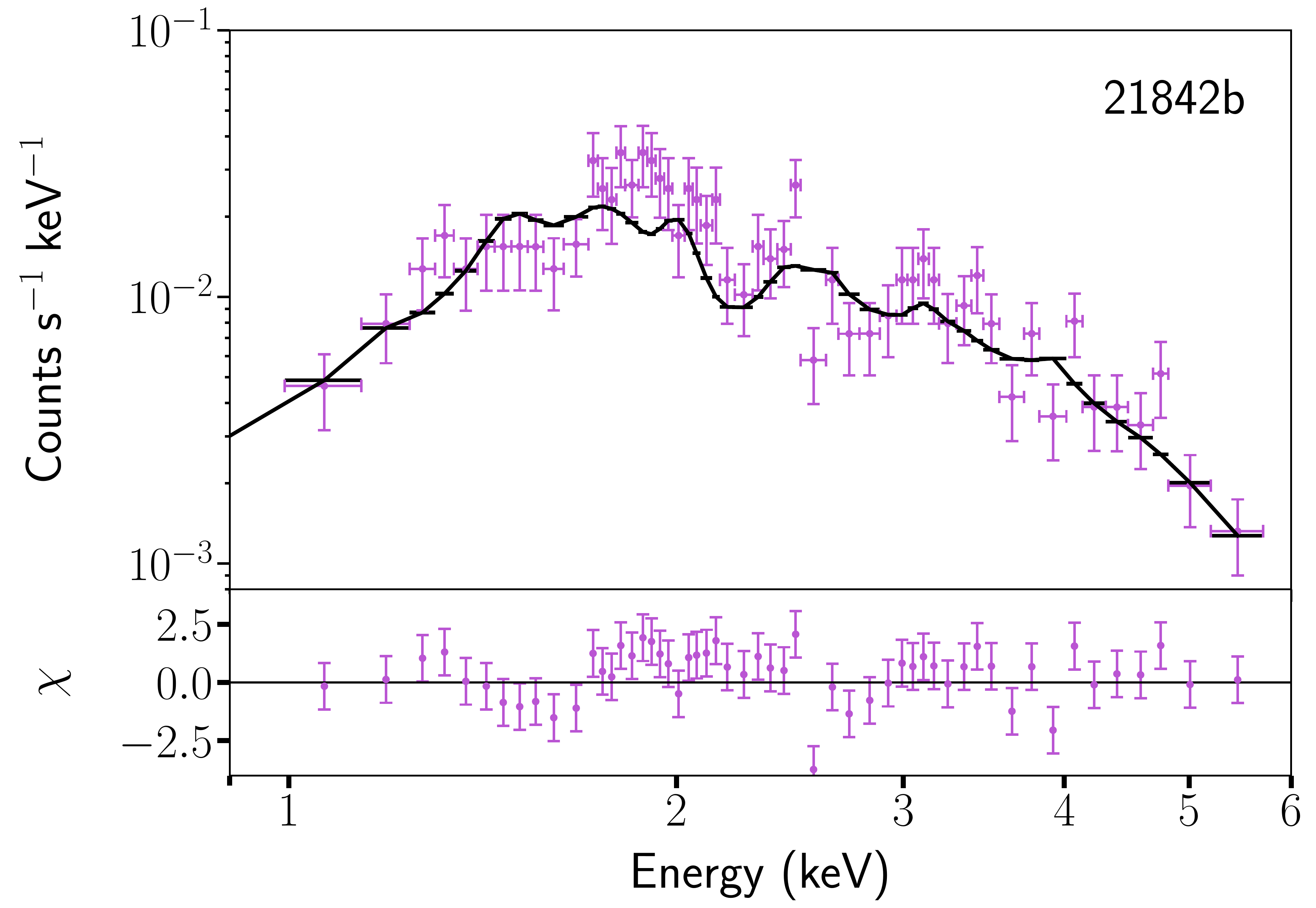}
        \includegraphics[width=0.32\textwidth]{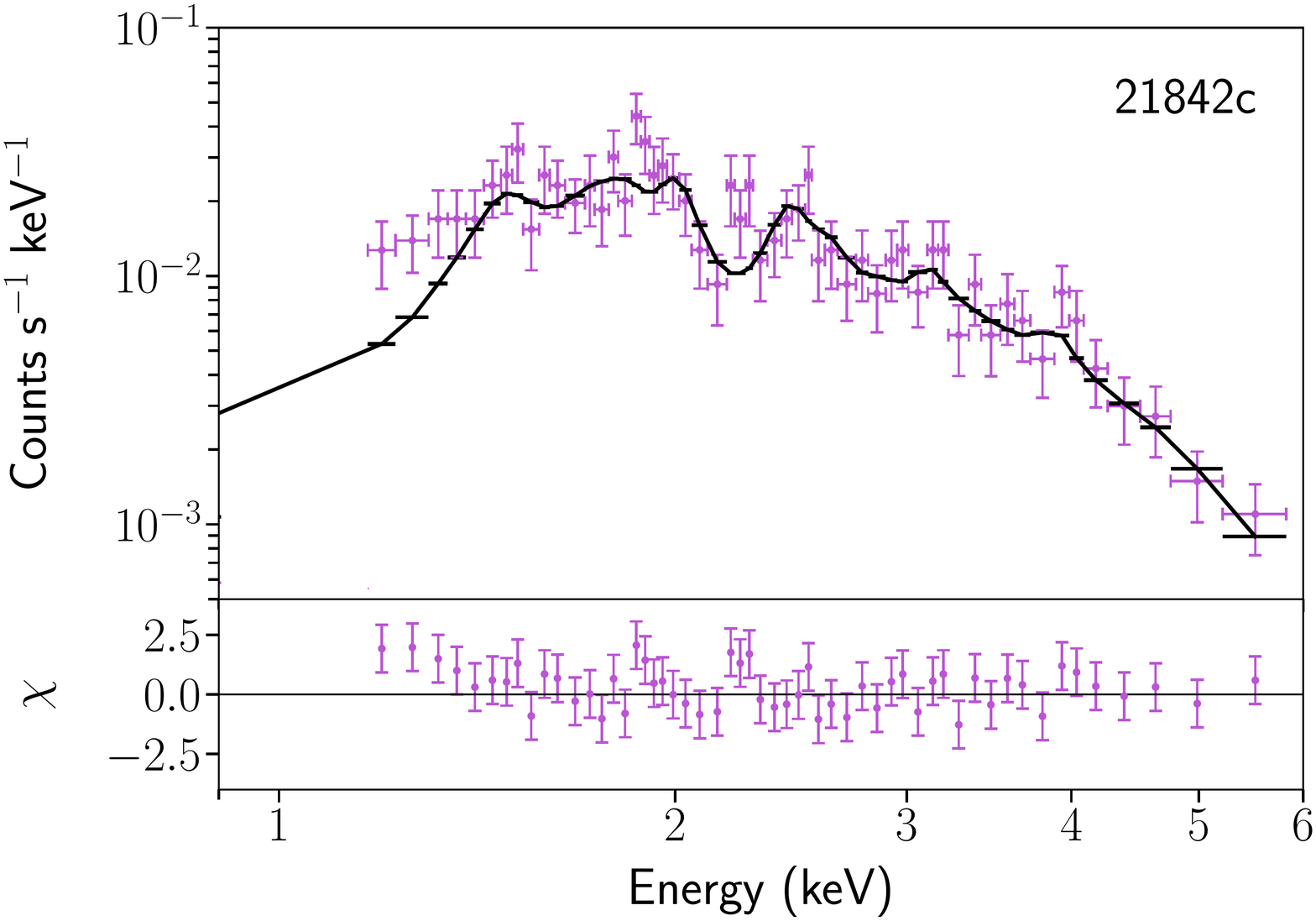}
            \includegraphics[width=0.32\textwidth]{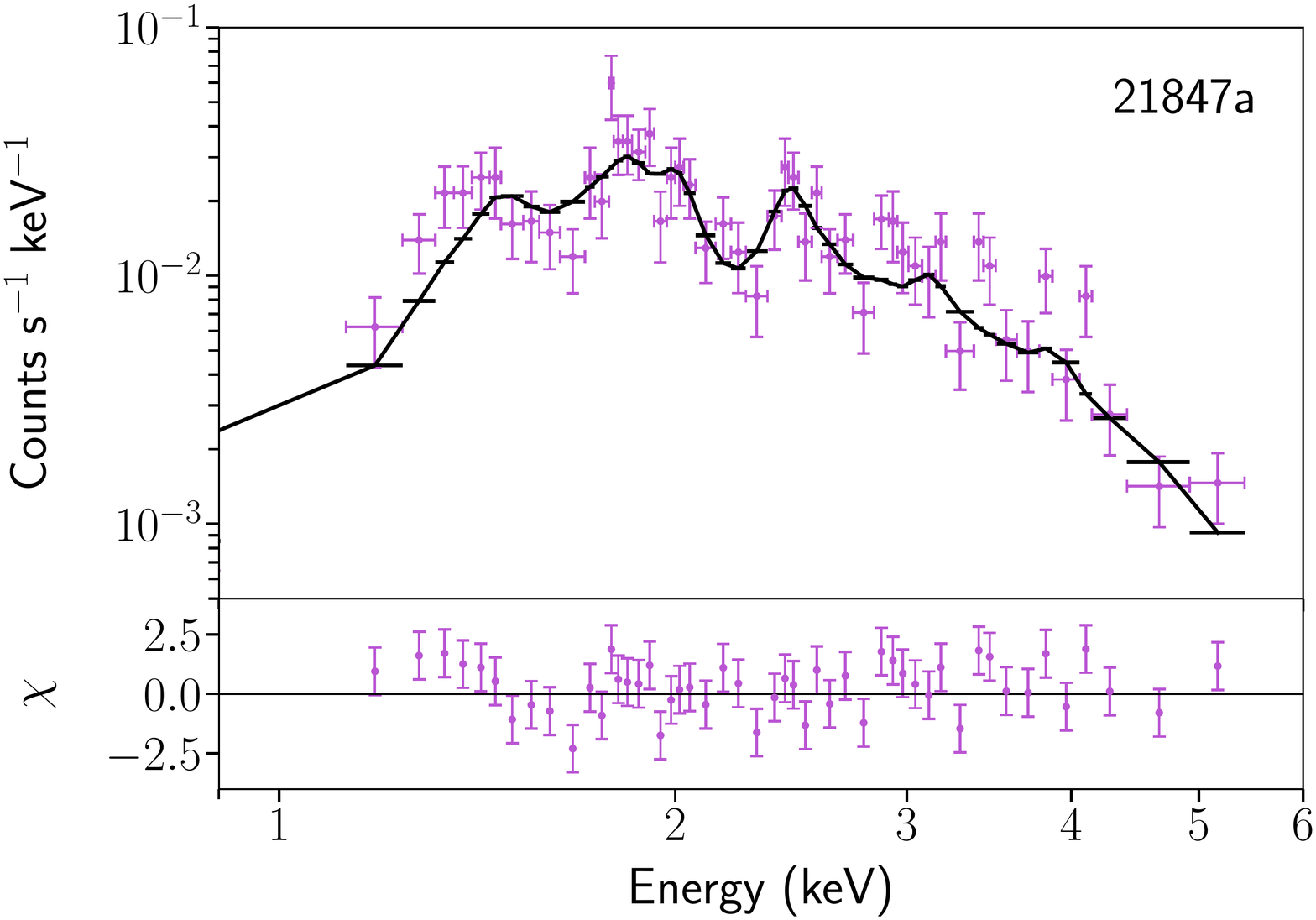}
\end{center}
\caption{Three examples of the spectra and residuals for our $\sim$15~ks bins. \textit{Left:} 21842b has the highest $\chi^2_{\rm red} = 71/51$ and temperature (kT) from our fits listed in \autoref{table:results}.  \textit{Center:} 21842c has the lowest $\chi^2_{\rm red} = 41/53$ in our sample. \textit{Right:} 21847a has the highest column density ($N_H$) where $N_H = 3.85 \times 10^{22}$ cm$^{-2}$ and the lowest temperature where $kT = 1.19$ keV from our $\approx$15~ks bins.}
\label{fig:spectra}
\end{figure*}

We use XSPEC Version 12.10.1f to background subtract our source spectra and model the background-subtracted spectra for each $\approx$15ks bin \citep{arnaud96}. We fit each spectrum separately using a model of \textsc{phabs}$\times$\textsc{apec}. This accounts for one absorption component \textsc{phabs} to represent both the galactic absorption and the absorption within the binary, and the emission from a hot, optically-thin plasma in collisional ionization equilibrium (CIE) using \textsc{apec}.  We set the abundance pattern to solar as measured by \cite{asplund09} and assume WR~20a has solar abundances. Example spectra are shown in \autoref{fig:spectra}.


Additionally, we evaluate the hardness ratio with finer phase resolution: specifically, we divide our $\approx$15~ks bins into three, evenly spaced bins of $\approx$5~ks.  We then extract the exposure corrected net counts in the medium energy band (M; $1-2$ keV) and the hard energy band (H; $2-7$ keV) for each bin. We calculate the hardness ratio as $HR2 = (H - M)/(H + M)$. We chose these energies to define our medium and hard energy bands to match those used in the models of \citet{montes13}.

\subsection{Optical and WR~20a Phase Determination} \label{sec:optdata}

To constrain the orbital period of WR~20a, we rely on ASAS-SN V- and g-band observations as well as observations from \textit{TESS}. ASAS-SN V-band observations were obtained between 2013 and 2018. ASAS-SN g-band observations were also obtained starting from 2018 through June 2021. The field of view of an ASAS-SN camera is 4.5 deg$^2$, the pixel scale is 8\farcs0 and the FWHM is typically $\approx2$ pixels. The ASAS-SN light curves were extracted as described in \citet{jayasinghe18} using image subtraction \citep{alard98,alard00} and aperture photometry on the subtracted images with a 2 pixel radius aperture. The AAVSO Photometric All-Sky Survey 
(APASS; \citealt{henden15}) was used for calibration. We corrected the zero point offsets between the different cameras as described in \citet{jayasinghe18}. The photometric errors were recalculated as described in \citet{jayasinghe19}.  We show these results in \autoref{fig:lightcurve}.

WR~20a was observed by the Transiting Exoplanet Survey Satellite (\textit{TESS}, \citealt{ricker14}) during Sectors 9 and 10 of the prime mission (between 2019-02-28 and 2019-04-22) and Sectors 36 and 37 of the extended mission (between 2021-03-07 and 2021-04-28).
Similar to the process applied to the ASAS-SN data, we used an image subtraction pipeline based on the ISIS package \citep{alard98,alard00} to obtain high fidelity light curves from the \textit{TESS} full frame images (FFIs).
For each sector we constructed a reference image from the first 100 FFIs of good quality obtained during that sector, excluding those with sky background levels or PSF widths above average for the sector.
A more in-depth discussion of this process and a detailed description of the \textit{TESS}-specific corrective procedures can be found in \citet{vallely19} and \citet{vallely21}.
During the prime mission \textit{TESS} FFIs were obtained at a 30-minute cadence; this was shortened to a 10-minute cadence during the extended mission.
Since we have data from two sectors in the prime mission and two sectors in the extended mission, 
we have more 10-minute images but the total time covered by 30-minute and 10-minute images is equal.
We thus have over 9,000 \textit{TESS} observations of WR~20a which we use in our analysis here.

This pipeline produces excellent differential flux light curves, but the large pixel scale of \textit{TESS} makes it difficult to obtain reliable measurements of the reference flux directly.
As such, here we follow the example set by \cite{jayasinghe20} and estimate the reference flux using the \verb"ticgen" software package \citep{ticgen,2018TIC}.
We convert the \verb"ticgen" magnitude estimates into fluxes using an instrumental zero point of 20.44 electrons per second in the FFIs, based on the values provided in the \textit{TESS} Instrument Handbook \citep{TESSHandbook}.
Flux is then added to the raw differential light curves such that the median of each sector's observations matches the estimated reference value.
A minor multiplicative scaling factor (typically around 15\%) is also included to ensure that the observed variable oscillation strength between the four sectors is appropriately matched.
This allowed us to produce the normalized flux light curves that we present in this work.

We present the first \textit{TESS} light curve for WR~20a in this paper in \autoref{fig:lightcurve}. We were unable to use the \textit{TESS} optical light curve to fit a contact binary model using \textsc{phoebe} version 2.3 in order to determine the binary parameters \citep{prsa16,conroy20}. 
The fits we obtain using \textsc{phoebe} do not fit the light curve of WR~20a well due to the Wolf-Rayet stars in this binary since \textsc{phoebe} is designed to fit stars with defined boundaries. 
As a result we use the binary parameters from \cite{bonanos04} who used the Warsaw telescope at Las Campanas Observatory, Chile and reduced using the OGLE-III pipeline \citep{udalski03}, when necessary to understand the WCR and the possible fate of the massive stars.

\section{Results} \label{sec:results}

We fit the spectra of our $\approx$15~ks bins as described in \autoref{sec:xraydata} to find the column density ($N_{\rm H}$) obscuring WR~20a, and temperature of the plasma producing the emission, $kT$. We report the results of the fits for each bin in \autoref{table:results} along with the corresponding orbital phase, the net counts, the normalization factor of the \textsc{apec} model, and the reduced $\chi^2$ defined as $\chi^2$/d.o.f. (where d.o.f is the degrees of freedom) associated with each bin. Our best-fit models for each of the 24 bins have parameter ranges of $2.26\times10^{22}$~cm$^{-2}<N_{\rm H}<3.85\times10^{22}$cm$^{-2}$ and 1.19\,$<kT<$\,2.03~keV.

We note that in some of our bins, including the left and middle panels in \autoref{fig:spectra}, there are large residuals around the Si\textsc{xiii} line at $\approx$1.85~keV. This Si line could indicate super-solar Si abundances in the winds of WR~20a. However when we fit the spectra with a model allowing for variable Si abundances, using a \textsc{phabs}$\times$\textsc{vapec} model, the uncertainties in the best-fit results were still consistent with solar values of the Si abundance. As such, we kept Si frozen to the solar value rather than thawing it and allowing it to vary.

\begin{figure*}
\begin{center}
    \includegraphics[width=\textwidth]{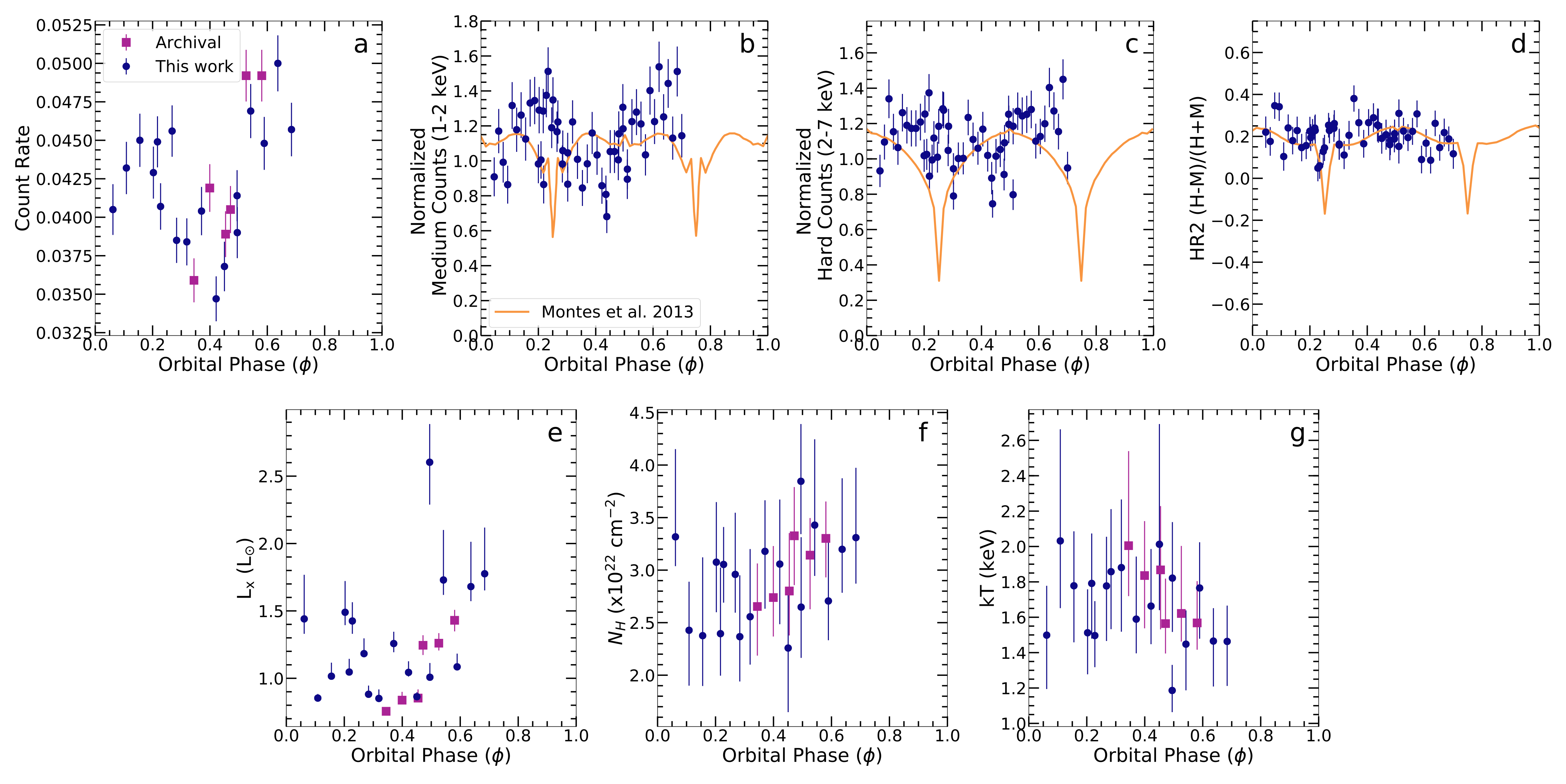}
\end{center}
\caption{
    \textit{Top:} Observations from archival data are shown as square points and new data in this work are shown as circles.  Models shown in panels $b$, $c$, and $d$ are from \cite{montes13}. \textit{a:} Count rate from 15~ks bins. \textit{b:} Normalized medium counts from the 1$-$2 keV range from 5ks bins. \textit{c:} Normalized hard counts from the 2$-$7 keV range from 5ks bins. \textit{d:} Hardness ratio $(H-M)/(H+M)$ for 5ks bins. \textit{Bottom:} Parameters derived from spectra. \textit{e:} $L_{\rm X}$ derived from the 15~ks bins. \textit{f:} Column density $N_{\rm H}$ derived from 15~ks bins. \textit{g:} kT derived from 15~ks bins.}
\label{fig:panels}
\end{figure*}

We plot the $N_{\rm H}$ and $kT$ values for the 24 bins in panels $f$ and $g$ of \autoref{fig:panels}. The square points show the archival observations that were first presented in \cite{naze08}, and the circular points are the new data from this work. Neither $N_{\rm H}$ nor $kT$ has a clear trend as a function of orbital phase ($\phi$). We also plot the count rate for each bin as a function of phase in panel $a$ of \autoref{fig:panels}. Unlike the other quantities, the count rate does display a coherent trend, increasing to a local maximum at $\phi\approx$0.2, decreasing to a minimum at $\phi\approx$0.4, and then peaking at a global maximum at $\phi\approx0.6-0.7$.  This indicates the X-ray emission is variable with the orbital phase of the binary, a feature not observed in the optical light curve shown in \autoref{fig:lightcurve}. Finally, we calculate the emitted (unabsorbed) X-ray luminosity for the $0.5-7$~keV range, $L_{\rm X}$, of WR~20a as a function of phase and plot that in panel $e$ of \autoref{fig:panels}.  $L_{\rm X}$ does not display a coherent trend as a function of phase.  We note that the highest luminosity point of $L_{\rm X}\approx2.7_{-0.31}^{+0.28}~L_\odot$ has the greatest $N_{\rm H}$ of our bins, with $N_{\rm H}=(3.85_{-0.50}^{+0.54})\times10^{22}$~cm$^{-2}$.

We calculate the hardness ratio (HR2) using our smaller $\approx$5~ks bins as described in \autoref{sec:xraydata}.  The hardness ratio is made up of medium counts ($1-2$~keV) and hard counts ($2-7$~keV), and is calculated as $HR2 = (H - M)/(H + M)$. We plot the net medium counts, net hard counts, and the hardness ratio as a function of phase in panels $b$, $c$, and $d$ of \autoref{fig:panels}. 
The counts in both the medium and hard bands appear to be following a similar trend with dips at $\phi\approx$0.45.  HR2 appears to be constant over the observed orbital phases, from $\phi = 0.0-0.7$.  We overplot the model from \cite{montes13} in panels $b$, $c$, and $d$ of \autoref{fig:panels} which predicted the X-ray emission from the WCR of WR~20a.  These models were normalized to the data, but notably do not fit the shape of our observations.

We averaged the $0.5-7$~keV X-ray count rate in bins of 0.25$\phi$ centered around $\phi$ = 0.125, 0.375, and 0.625 to capture the peaks and trough we observe in the X-ray light curve.  We converted to fluxes using WebPIMMS and the thermal Bremsstrahlung model with a kT of 1.7~keV and a galactic $N_{\rm H}$ of 3.0$\times$10$^{22}$ cm$^{-2}$ which we used because they are the average values of $kT$ and $N_{\rm H}$ from our bins. We then plotted these fluxes as yellow diamonds on \autoref{fig:lightcurve}. We measured the difference in the value of the two peaks as 2.5$\sigma$. 

\begin{deluxetable*}{lcccccc}
\tablecolumns{7}
\tablewidth{0pt} \tablecaption{Results \label{table:results}} 
\tablehead{\colhead{ObsID} & \colhead{Orbital Phase} & \colhead{Net Counts} & \colhead{$N_{\rm H}$ ($\times$10$^{22}$ cm$^{-2}$)} & \colhead{kT (keV)} & \colhead{norm (cm$^{-5}$)} & \colhead{$\chi^2$/d.o.f.} }  
\startdata
6410a & 0.344 & 692 & 2.65$_{-0.47}^{+0.41}$ & 2.00$_{-0.29}^{+0.53}$ & 1.14e-3 & 46/51 \\
6410b & 0.399 & 743 & 2.74$_{-0.37}^{+0.49}$ & 1.84$_{-0.30}^{+0.31}$ & 1.26e-3 & 73/60 \\
6410c & 0.454 & 738 & 2.80$_{-0.42}^{+0.55}$ & 1.87$_{-0.34}^{+0.36}$ & 1.29e-3 & 61/54 \\
6411a & 0.472 & 770 & 3.33$_{-0.47}^{+0.46}$ & 1.56$_{-0.17}^{+0.25}$ & 1.77e-3 & 62/55 \\
6411b & 0.526 & 862 & 3.14$_{-0.52}^{+0.35}$ & 1.62$_{-0.16}^{+0.38}$ & 1.83e-3 & 74/66 \\
6411c & 0.581 & 913 & 3.30$_{-0.37}^{+0.35}$ & 1.57$_{-0.15}^{+0.24}$ & 2.03e-3 & 88/66 \\
20133a & 0.451 & 582 & 2.26$_{-0.61}^{+0.57}$ & 2.01$_{-0.37}^{+0.68}$ & 1.31e-3 & 40/43 \\
20133b & 0.495 & 616 & 2.65$_{-0.48}^{+0.67}$ & 1.82$_{-0.30}^{+0.32}$ & 1.52e-3 & 39/46 \\
20134a & 0.228 & 801 & 3.05$\pm$0.36          & 1.50$_{-0.18}^{+0.19}$ & 1.98e-3 & 58/57 \\
20134b & 0.284 & 739 & 2.37$_{-0.43}^{+0.58}$ & 1.86$_{-0.32}^{+0.35}$ & 1.33e-3 & 42/53 \\
21842a & 0.062 & 654 & 3.32$_{-0.28}^{+0.83}$ & 1.50$_{-0.30}^{+0.28}$ & 2.00e-3 & 67/49 \\
21842b & 0.109 & 650 & 2.43$_{-0.53}^{+0.46}$ & 2.03$_{-0.38}^{+0.63}$ & 1.29e-3 & 71/51 \\
21842c & 0.156 & 679 & 2.38$_{-0.48}^{+0.75}$ & 1.78$_{-0.32}^{+0.31}$ & 1.53e-3 & 41/53 \\
21842d & 0.203 & 720 & 3.08$_{-0.48}^{+0.57}$ & 1.51$_{-0.23}^{+0.24}$ & 2.10e-3 & 54/50 \\
21847a & 0.494 & 683 & 3.85$_{-0.50}^{+0.54}$ & 1.19$_{-0.12}^{+0.14}$ & 2.94e-3 & 62/48 \\
21847b & 0.542 & 718 & 3.43$_{-0.48}^{+0.82}$ & 1.45$_{-0.26}^{+0.19}$ & 2.36e-3 & 55/57 \\
21847c & 0.589 & 681 & 2.71$_{-0.37}^{+0.59}$ & 1.77$_{-0.29}^{+0.26}$ & 1.63e-3 & 47/54 \\
21847d & 0.637 & 764 & 3.20$_{-0.41}^{+0.68}$ & 1.47$_{-0.26}^{+0.18}$ & 2.31e-3 & 72/57 \\
21847e & 0.684 & 768 & 3.31$_{-0.44}^{+0.66}$ & 1.46$_{-0.25}^{+0.20}$ & 2.43e-3 & 48/56 \\
21848a & 0.217 & 798 & 2.40$_{-0.40}^{+0.63}$ & 1.79$_{-0.31}^{+0.28}$ & 1.58e-3 & 72/59 \\
21848b & 0.268 & 753 & 2.96$_{-0.36}^{+0.59}$ & 1.78$_{-0.31}^{+0.28}$ & 1.78e-3 & 59/58 \\
21848c & 0.319 & 630 & 2.56$_{-0.46}^{+0.64}$ & 1.88$_{-0.36}^{+0.39}$ & 1.28e-3 & 52/49 \\
21848d & 0.371 & 673 & 3.18$_{-0.55}^{+0.49}$ & 1.59$_{-0.19}^{+0.35}$ & 1.80e-3 & 49/52 \\
21848e & 0.422 & 628 & 3.06$_{-0.57}^{+0.61}$ & 1.66$_{-0.22}^{+0.32}$ & 1.53e-3 & 58/45 
\enddata
\end{deluxetable*}


\section{Discussion} \label{sec:discuss}

\subsection{X-Ray Light Curve Asymmetry} \label{sec:wcr}
We measured the amplitude and orbital phase of the peaks of the X-ray light curve and found that they are asymmetric. In the X-ray light curve, the peak at $\phi$=0.625 is 2.5$\sigma$ higher than the peak at $\phi$=0.125, and shifted  by $\phi\approx$0.1 with respect to the optical light curve. This offset
likely is the result of some asymmetry in the X-ray emitting region. We explore some of the possibilities for producing an asymmetric X-ray light curve below, but to constrain the significance of this asymmetry we need additional X-ray data with finer time resolution and greater phase coverage.

Asymmetric light curves produced by the X-rays emanating from the WCRs of massive binary systems are not uncommon \citep[e.g.,][]{lomax15,pollock18, garofali19, pollock21, pradhan21}. Most of these systems typically consist of two stars of different masses, usually an evolved WR star and O-star companion. As a result, the WCR is powered by two sources with different mass-loss rates, wind velocities, and chemical compositions \citep{vink21}. However, WR~20a is unique in that it is a close-in ($P=3.686$~day), tidally locked, massive binary with a circular orbit ($e=0$) consisting of two nearly identical mass ($M_{\rm 1} \approx 83$ and $M_{\rm 2} \approx 82\; M_{\odot}$), hydrogen-rich Wolf-Rayet stars that are likely in a pre-LBV phase \citep{bonanos04, rauw05, smith08}. 

Given its short orbital period and non-eccentric orbit, \citet{bonanos04} found that the orbital separation between these two stars is $\approx 51$~$R_{\rm \odot}$ (assuming an inclination angle of $74.5^{\circ}$ which was determined from the light curve) and that each component has a mean radius of $19.6 \; \rm R_{\odot}$ \citep{rauw05}. Both stars are still within their Roche lobes, therefore the binary is detached and not currently undergoing Roche lobe overflow mass transfer \citep{rauw05}. However, since the stars are tidally locked, they are in synchronous rotation. This leads to fast rotational velocities and tidal forces between the two stars, which causes the stars to deviate from a spherical structure. Instead they are oblate along their equator and flattened at the poles. This tidal and rotational distortion leads to a non-uniform luminosity and effective temperature across each star \citep{maeder10}. This results in non-uniform line-driven winds that are stronger at the poles as compared to the equator, potentially causing the overall mass-loss to be bipolar \citep{bonanos04, maeder10, meyer21, vink21}. Additionally, the fast rotation leads to rotational mixing (i.e., chemical homogeneous evolution), which agrees with the enhanced nitrogen and depleted carbon abundances in the observed spectra of WR~20a \citep{rauw05, deMink2009b, martins13, song16}. 

Asymmetric mass-loss will affect the structure of the WCR between these two stars and may lead to Kelvin-Helmholtz instabilities at the WCR interface that can enhance the density. This will lead to a greater absorption of the X-rays produced in the WCR \citep{lamberts12, naze18}. \cite{rauw05} assumed the mass-loss rates for the two stars in WR~20a were identical, however using the multiple epochs of spectra in \cite{rauw05} along with the orbital parameters determined in \cite{bonanos04} one could separate the spectra of the individual stars in this system using Doppler tomography \citep{bagnuolo91,gies04,massey12}. This separation of the individual stellar components of WR~20a would allow for cleaner spectral typing and the ability to determine the differences in the stellar wind properties between the two component stars. Identifying the mass-loss rates of the individual stellar components in WR~20a is a necessary step to understanding the asymmetric X-ray light curve.

Additionally, the WCR is not planar as previous analytical and numerical modeling by \citet{antokhin04} and \citet{montes13} assumed, as the Coriolis forces caused by the binary's orbital motion warps the edges of the WCR into a spiral \citep{rauw04, lamberts12, lomax15}. As such, we suggest that the asymmetric mass-loss for the components of WR~20a and the warped non-planer structure of the WCR is likely responsible for the asymmetric X-ray light curve we observe with \textit{Chandra}. In addition, hydrodynamical instabilities that develop in the WCR may create gas clumping that could increase the absorption of X-ray photons and thus further affecting the X-ray light curve. In agreement, numerous observations and modeling of winds from evolved fast-rotating massive stars find that their winds should be axisymmetric. This will affect the structure of the WCR and wind bubbles that they produce and thus affect the stars' surrounding circumstellar medium and the resulting supernova explosion 
\citep{eldridge07}. Regardless, future numerical work exploring the X-ray emission from WCRs produced by a massive binary should take asymmetric wind mass-loss rates, orbital motion, and systemic motion into account to compare with the work presented here.

\subsection{WR~20a: A Gravitational Wave Source in our Backyard?} \label{sec:ligo}

The two stars in WR~20a are massive enough to produce a black hole from their core-collapse supernova explosion, leaving behind a binary black hole system similar to those observed in gravitational wave events detected by LIGO and LISA \citep{hainich18}. As such, we studied the properties of the binary black hole system that WR~20a might produce in order to create a better understanding of the possible progenitor landscape involved in binary black hole mergers.

Since the stars in WR~20a are most likely tidally locked to the binary’s orbital period, and have a $R_*\approx 19R_{\odot}$ radius and $M_*\approx 80M_{\odot}$ mass, they would be rotating at $\approx 29\%$ of their breakup velocity, $\Omega_{\rm break} = (GM_{*}/R_{*}^3)^{1/2}$. As such they could experience enhanced rotational mixing and a chemically homogeneous evolution \citep{Yoon2005}. Furthermore, as proposed by \citet{mandel16}, the evolution of an isolated close stellar binary whose stars are tidally locked and retain large rotation rates throughout their entire life, may lead to the production of a black hole binary. In this scenario the rotationally induced chemically homogeneous evolution prevents the build-up of an internal chemical gradient causing the stars to remain compact as they evolve, eventually leading to the formation of a compact black hole binary. However, if these stars assemble helium cores more massive than $\approx 30 M_{\odot}$, they will eventually produce electron-positron pairs at regions with high entropy and temperature, leading to the so called ``pair-instability’’ \citep{Fowler1964}. Once this happens the star becomes unstable and can experience a single violent pulse that disrupts the entire star (pair instability supernova), or undergo intense pulsations that induce mass loss and lead to a less massive star that suffers a normal core collapse (pulsational pair-instability) that could lead to a supernova explosion and the formation of a massive black hole \citep{Woosley2017}. 

Thus, if the $\approx 80M_{\odot}$ stars in WR~20a retain a significant fraction of their mass, they will form a massive helium core and eventually go through the ``pair-instability’’ phase, making their evolutionary paths very uncertain. As noted in the paper by \cite{Woosley2017}, the occurrence of a pulsational pair-instability (PPI) phase in massive stars depends on their metallicities and rotation rates. Metallicity plays an important role in determining the star’s mass loss rate and the mass of its helium core, which in turn determines if there will be a PPI phase. In order to retain helium cores massive enough to enter the PPI phase, models with solar metallicity require a significant reduction in their mass loss rates. Moreover, adding rapid rotation to the stellar models leads to chemical mixing which increases the mass of the helium core for a given main sequence mass, reducing the stellar mass threshold for PPI to occur. In the work by \cite{Woosley2017}, $80M_{\odot}$ non rotating solar metallicity stellar models that go through the PPI phase yield PreSN masses ranging from 40 to $56M_{\odot}$ depending on the mass loss rate. Meanwhile, rapidly rotating sub-solar metallicity models with $80M_{\odot}$ that go through the PPI phase yield PreSN masses ranging from 30 to $59M_{\odot}$, depending on the mass loss rate.  

In the following calculations we will assume that the stars in WR~20a undergo a PPI phase, described by the rapidly rotating model C80B with $80M_{\odot}$ and sub solar metallicity, which experienced pulsations for $\approx 45$ days leading to substantial mass loss and yielding a PreSN mass of $44.88M_{\odot}$. However, since the stars in WR~20a seem to have solar composition, they might suffer from further mass loss that prevents them from entering the PPI phase yielding PreSN stars with masses $M\lesssim 30M_{\odot}$ and ultimately leading to less massive BHs. Assuming that the stars in WR~20a evolve in a similar fashion as the model C80B from \cite{Woosley2017}, we expect them to end their lives with $\approx 45M_{\odot}$ and a fraction of the angular momentum content they currently have. Moreover, the final orbital parameters of WR~20a will be determined by a competition between the mass loss from the PPI (widening the orbit), and the interaction between this slowly ejected material and the stars (tightening the orbit through drag forces). This hinders an accurate estimation of the merger time scale through gravitational waves for WR~20a, which would be easily over 50 Gyrs just considering the separation and eccentricity expected from an instantaneous mass loss of $\approx70M_{\odot}$ from the two stars.

\begin{figure*}

        \includegraphics[trim={0.9cm 0.0cm 1.0cm 0.0cm},width=0.335\linewidth]{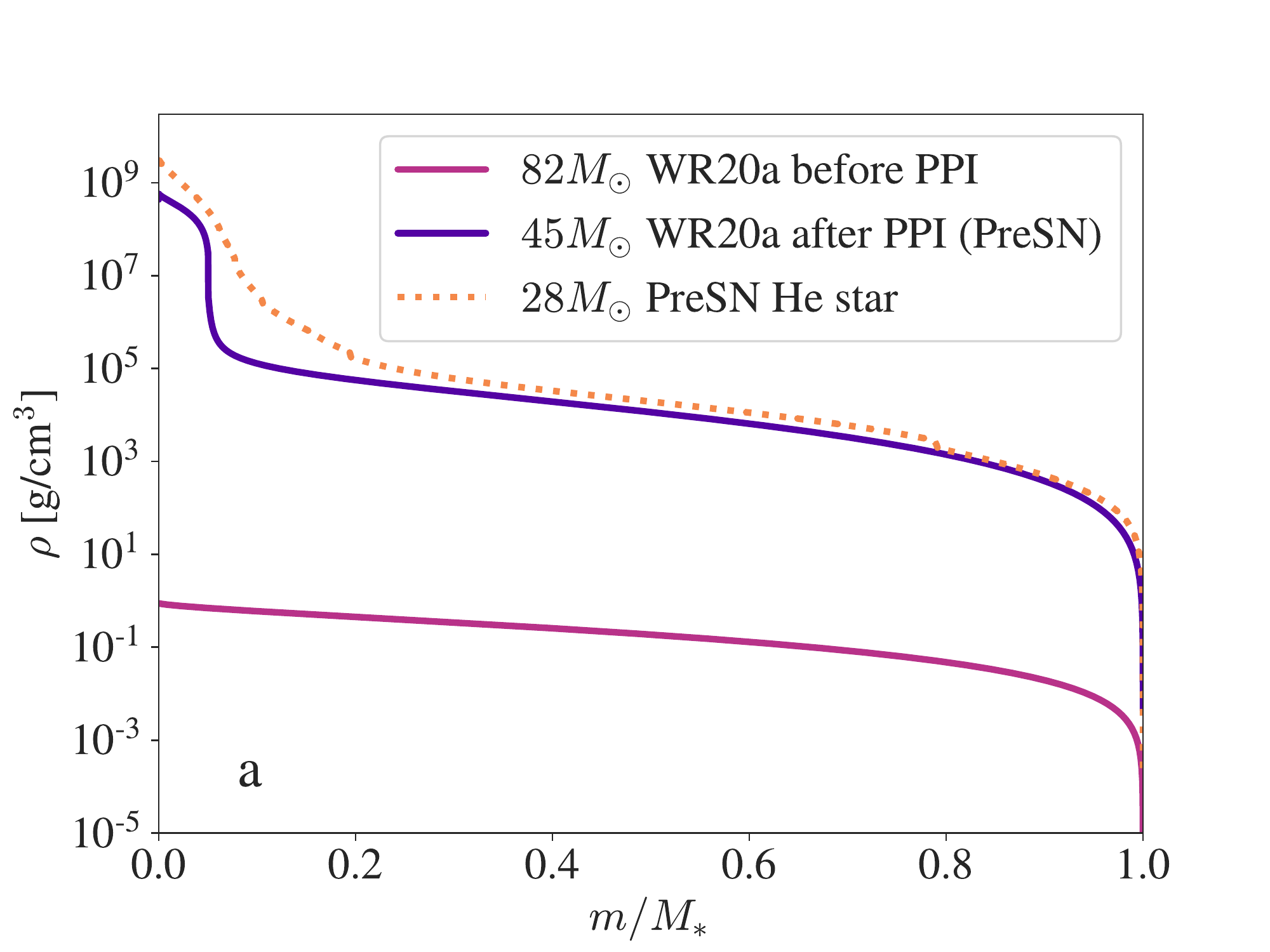} 
        \includegraphics[trim={0.8cm 0.0cm 1.1cm 0.0cm},width=0.335\linewidth]{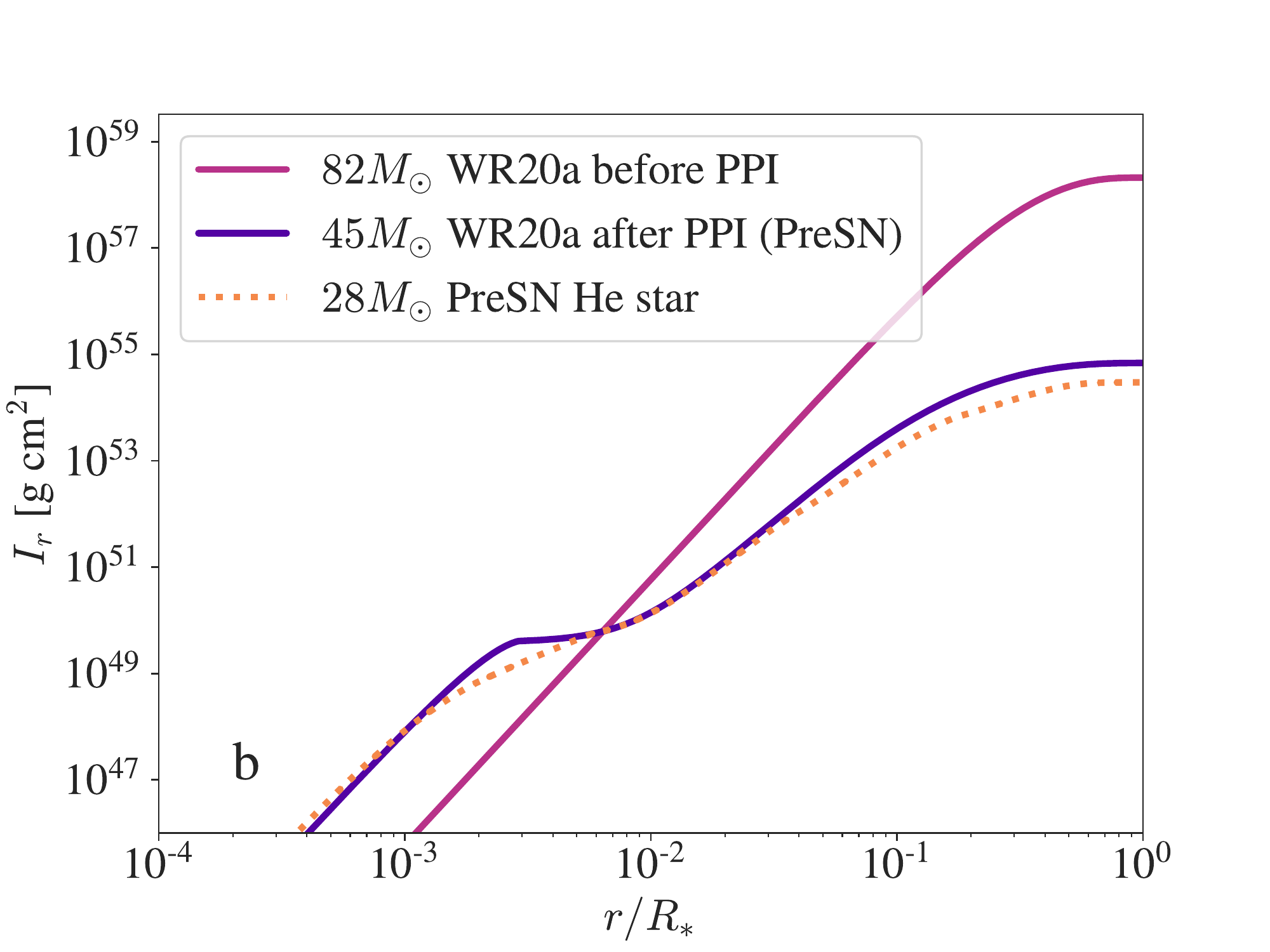} 
        \includegraphics[trim={0.8cm 0.0cm 1.1cm 0.0cm},width=0.335\linewidth]{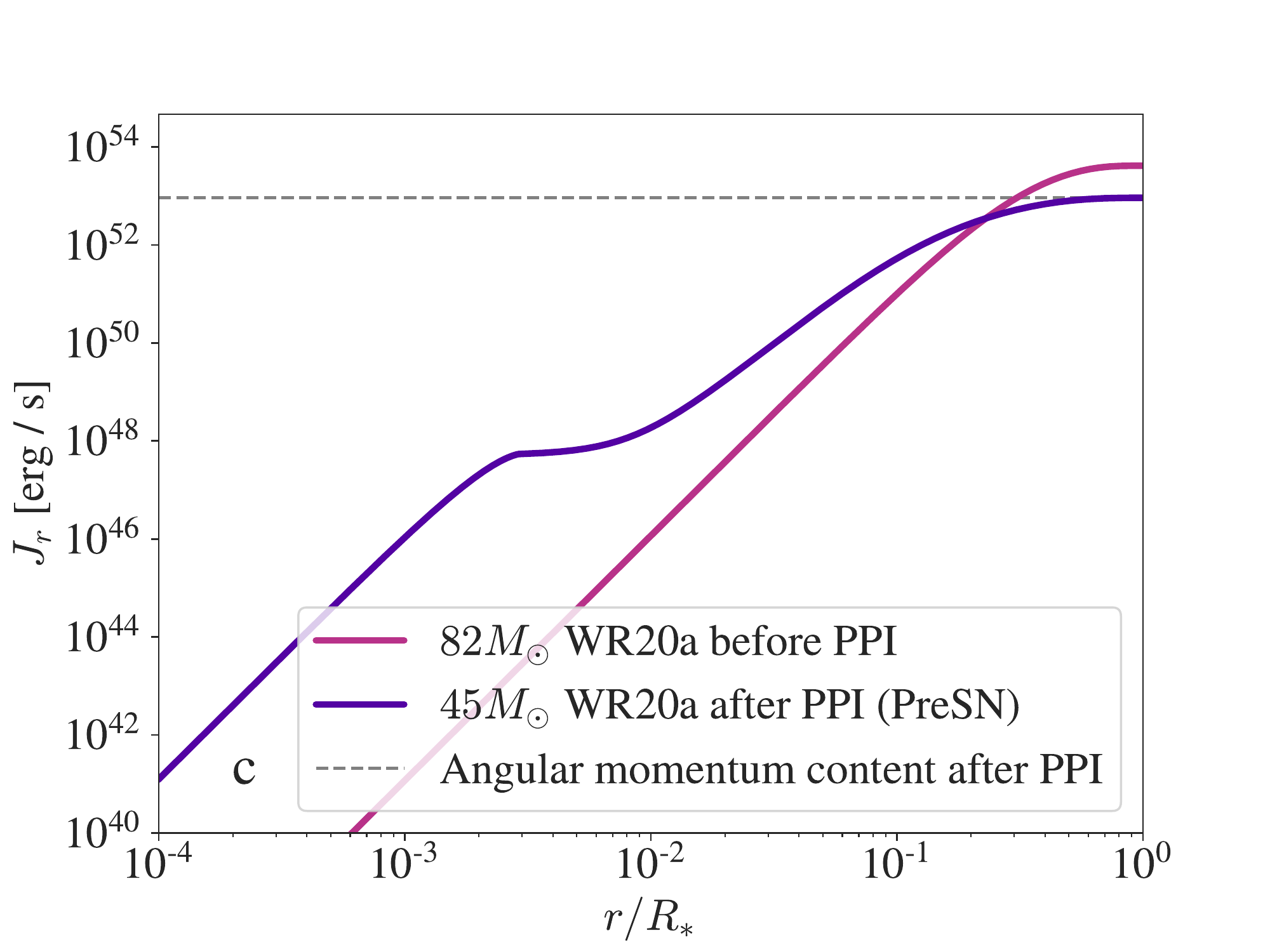}

        \includegraphics[trim={1.0cm 0.5cm 4.1cm 0.0cm},width=\linewidth]{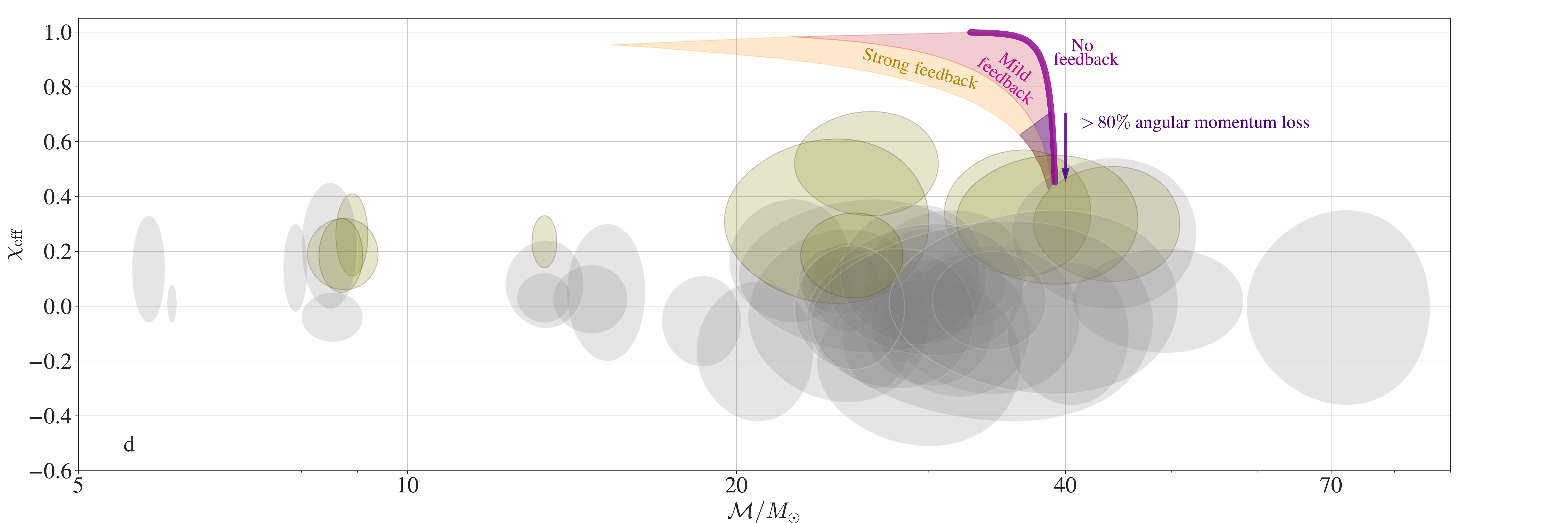} 
        
    \caption{Top panels: Mass and angular momentum distribution of stars in WR~20a. a) Density as a function of the normalized coordinate mass $m/M_{*}$ for a $82M_{\odot}$ polytropic star with $\gamma=4/3$ (radiation dominated), a PreSN 28 $M_{\odot}$ He star from WH06, and a $45M_{\odot}$ Nested polytropic star. b) Moment of inertia as a function of the normalized radius $r/R_{*}$ for the same three stellar models. c) Angular momentum content as a function of the normalized radius $r/R_{*}$ of the $82M_{\odot}$ polytropic star (assuming tidal synchronization), and the $45M_{\odot}$ star (assuming angular momentum conservation after the PPI phase). Bottom panel: d) Effective spin parameter $\chi_{\rm eff}$ and chirp mass $\mathcal{M}$ of the 44 BBH LIGO events reported in the GWTC-2 Catalogue \citep{O2_LIGO2021}, together with the expected range of effective spins and chirp masses of WR~20a, assuming a $45M_{\odot}$ nested polytropic star and different amounts of angular momentum conservation after the PPI phase.}
    
    \label{fig:LIGOplot}
\end{figure*}

In spite of the uncertainty in the final orbital parameters of WR~20a, we can still obtain some information on the expected properties of the black hole binary that could be formed from such a system. In order to do this we start by assuming the two tidally locked stars in WR~20a have masses equal to $82M_{\odot}$, are radiation dominated and described by a $\gamma=4/3$ polytrope. Their mass, moment of inertia and angular momentum distributions are shown in the top panels of \autoref{fig:LIGOplot}. Moreover, if the stars undergo PPI and lose $37M_{\odot}$ during their pulsations, the remaining inner $45M_{\odot}$ should continue to evolve past the ``pair instability’’ phase as a less massive He star with a reduced or negligible H envelope. 

Regardless of the specific evolution followed by these remaining $45M_{\odot}$ stars, they must end their life with the formation of a dense compact core composed mostly of heavy elements (Fe, Si, S, etc.) surrounded by an extended envelope with abundant lighter elements.  Thus, we used a nested polytropic star composed of a $2.25M_{\odot}$ core and a $42.75M_{\odot}$ envelope with $R_*=0.9R_{\odot}$ (described by  $\gamma_{\rm core} = 1.83$ and $\gamma_{\rm env}=1.25$\, respectively) to mimic the structure of the $45M_{\odot}$ star just before its collapse. The top panels of \autoref{fig:LIGOplot}, show the mass and moment of inertia distributions of the $45M_{\odot}$ nested polytropic star along with a $28M_{\odot}$ PreSN helium star from \cite{Woosley2006} and the $82M_{\odot}$ $\gamma=4/3$ polytropic star.

Assuming that the PPI phase only removed the angular momentum contained in the ejected $37M_{\odot}$ of each star, we can obtain an upper limit to the angular momentum content of the stars before their collapse into BHs. The rightmost top panel from \autoref{fig:LIGOplot} shows the angular momentum distribution of the $82M_{\odot}$ tidally locked star along with the angular momentum content of the star after the PPI phase (dashed line), and the corresponding angular momentum distribution of the PreSN $45M_{\odot}$ star assuming rigid body rotation. The structural information of the $45M_{\odot}$ star allows us to use the method described in \cite{Batta2019} to compute the effective spin parameter $\chi_{\rm eff}$ and the chirp mass $\mathcal{M}$ of the binary black holes (BBH) formed from the collapse of the $45M_{\odot}$ stars described in the top panels of \autoref{fig:LIGOplot}:
\begin{equation}
    \chi_{\rm eff} = \frac{m_1 S_1 + m_2 S_2}{m_1+m_2}\cdot \bar{J},\qquad
    \mathcal{M} = \frac{(m_1m_2)^{3/5}}{(m_1+m_2)^{1/5}},
\end{equation}
which are defined by the individual BH's masses ($m_1$ and $m_2$) and spins ($S_1$ and $S_2$), and the binary's normalized angular momentum $\bar{J}$. 

The effective spin is a measure of the alignment between the sum of the weighted individual spins and the binary's normalized angular momentum $\bar{J}$. A positive $\chi_{\rm eff}$ implies that both BHs have their spin at least partially aligned with the binary's angular momentum (the angle between them is smaller than 90\textdegree). Meanwhile, a negative effective spin implies that one or both individual spins are anti-aligned with the binary's orbital angular momentum (the angle between them is larger than 90\textdegree). This parameter is important since it can contain information about the formation channel that formed the black hole binary \citep{Rodriguez_2016,Farr_2017,2020ApJ...894..129S}. A BBH formed from the evolution of isolated binary stars is mostly expected to have a positive effective spin, since mass transfer events and tidally locking work towards aligning the spin of the stars with the orbital angular momentum \citep{2018MNRAS.473.4174Z, 2018ApJ...862L...3S}. In such systems only a strong natal kick given to the BH during its formation could be able to alter such alignment. However, since the velocity imparted by these natal BH kicks is expected to be smaller than their orbital velocity \citep{Mandel_2016}, it is very difficult to form BBHs with negative effective spin from such isolated stellar binaries. Therefore, for these calculations we assumed that both stars have their spins aligned with the orbital angular momentum and that there is no natal kick during BH formation to alter this alignment. 

In order to explore scenarios where the stars lost more angular momentum during and after the PPI phase, we parameterized the angular momentum content of the $45M_{\odot}$ stars as done in \cite{Batta2019}. The results are shown in the bottom panel of \autoref{fig:LIGOplot}, where the purple solid line indicates the values of $\chi_{\rm eff}$ and $\mathcal{M}$ expected for BBHs formed without any accretion feedback and with different angular momentum contents. A larger angular momentum content leads to a larger effective spin, but also to a smaller chirp mass due to mass loss during the formation of rapidly rotating BHs. We also show scenarios with different intensities of accretion feedback (pink and orange shaded areas), which lead to significantly smaller BH masses and slightly reduced spins. When implementing mild accretion feedback, the BBH ends with slightly smaller $\chi_{\rm eff}$ and noticeable losses in chirp mass, as indicated by the pink shaded area. As the accretion feedback increases, the chirp mass and the effective spin parameter further decrease due to mass loss during BH formation, as shown by the orange shaded region.

If the $45M_{\odot}$ stars retain their angular momentum during and after the PPI phase, the BBHs would lie at the top of the pink and orange shaded areas with $\chi_{\rm eff}\gtrsim0.9 $ and $ 15\lesssim \mathcal{M}/M_{\odot} \lesssim 30$ depending on the intensity of the accretion feedback. Therefore, in order to produce BBHs with effective spins consistent with LIGO observations $\chi_{\rm eff}\lesssim 0.65$ \citep{O2_LIGO2021}, the stars in WR~20a should lose a large portion of their angular momentum content during or after the PPI phase. This is indicated by the purple shaded area that highlights the location of BBHs formed from the same $45M_{\odot}$ stars but with 80\% less angular momentum than the one retained after the PPI phase. 

We should keep in mind that the results shown in \autoref{fig:LIGOplot} only explore the properties of BBHs formed from a very specific evolution channel for WR 20a. If both stars have solar metallicity, they might lose enough mass to prevent the ``pair instability'' phase and lead to slightly smaller BH masses. Such mass and angular momentum loss would lead to smaller effective spin parameters and chirp masses than the ones obtained in our calculations. The loss of angular momentum in this scenario will be crucial in determining the properties of the BBH, since losing too much angular momentum will lead to a normal isolated binary evolution where one or more mass transfer events could occur, leading to the formation of a BBH \citep{Belczynski_2016}. But if these stars manage to retain enough angular momentum to evolve according to \citet{mandel16}, the BH binary would be significantly wider than in our calculations and it would not merge within a Hubble time scale as the observed LIGO events.

\section{Conclusions} \label{sec:conclusions}

We analyzed the light curve of the massive binary WR~20a using new X-ray observations from \textit{Chandra} and optical observations from \textit{TESS} and ASAS-SN to understand the time variability of the X-ray emitting wind collision region between two Wolf-Rayet stars and how it compares to the optical variation.  We consider the sources of asymmetry in the X-ray light curve and explore the possible fate of the two massive stars in this binary by exploring the possible future gravitational wave events this binary might produce.

\begin{itemize}
    \item We use the deepest \textit{Chandra} observations (268~ks) of the young open cluster Westerlund 2 to study the time variability of the X-rays that emanate from the WCR of the close-in Wolf Rayet binary WR~20a.
    \item We produce an X-ray light curve of this massive binary using $\approx$15~ks bins of \textit{Chandra} data and use spectral analysis with \textsc{ciao} to look for phase variability in the temperature (kT) or column ($N_H$) parameters.
    \item We present the first \textit{TESS} light curve for WR~20a. The fits we obtain using \textsc{phoebe} do not fit the light curve of WR~20a well due to the Wolf-Rayet stars in this binary since \textsc{phoebe} is designed to fit stars with defined boundaries.
    \item When we examined the X-ray light curve we found that there is asymmetry in the flux of the peaks. The flux of the peak at $\phi$=0.625 is larger than the peak at $\phi$=0.125 by 2.5$\sigma$. This asymmetry could be due to a number of reasons which include asymmetric wind mass loss rates from the two WR stars or an asymmetrical WCR due to effects like the Coriolis force and/or hydrodynamic instabilities that develop in the WCR.
    \item Finally, we look at the results of a BBH merger from this system to compare the angular momentum of the gravitational wave event to those observed in LIGO. We find that the angular momentum from the BBH WR~20a could produce is much higher than the majority of observed BBHs from LIGO.  This indicates that systems like WR~20a are likely not progenitors to the majority of BBH mergers observed in LIGO unless the system loses $\gtrsim 80$~\% of its angular momentum content.
\end{itemize}

In order to better understand the wind interactions in the WCR of WR~20a we need to observe the full period of the binary in X-rays. With \textit{Chandra} we have obtained observations of $\approx $2/3 of the period, and with future \textit{Chandra} observations of the final 1/3 of the orbit we can confirm the shifted X-ray light curve and the asymmetry in the peaks of the X-ray light curve. Additionally, longer baseline observations of the X-rays with precision timing with NICER would allow for a detailed understanding of the variability in the wind collision region. Finally, understanding the mass-loss rates from each individual star in this binary would allow for a better understanding of what we expect to see in the WCR of WR~20a. Using Doppler tomography to separate the spectra of the two WR stars presented in \cite{rauw05} would make it be possible to model the mass-loss rates from each star individually rather than assuming equal mass-loss rates based on their spectral types, which is not guaranteed for WR stars.

\begin{acknowledgements}

We would like to thank Ilya Mandel, Ryosuke Hirai, Kris Stanek, Todd Thompson, and the OSU Galaxy/ISM Group Meeting for providing insightful conversations during the preparation of this work. Support for this work was provided by the National Aeronautics and Space Administration through Chandra Award Number GO8-19003X issued by the Chandra X-ray Center, which is operated by the Smithsonian Astrophysical Observatory for and on behalf of the National Aeronautics Space Administration under contract NAS8-03060. G.M.O. was supported by the Ohio State University through the Distinguished University Fellowship as well as George P. and Cynthia Woods Mitchell Institute for Fundamental Physics and Astronomy at Texas A\&M University. L.A.L. acknowledges support by the Simons Foundation, the Heising-Simons Foundation, and a Cottrell Scholar Award from the Research Corporation for Science Advancement. A.L.R. acknowledges support from the National Science Foundation (NSF) Astronomy and Astrophysics Postdoctoral Fellowship under award AST-2202249. K.F.N acknowledges support by NASA through the NASA Hubble Fellowship grant HST-HF2-51516. Support for TJ was provided by NASA through the NASA Hubble Fellowship grant HF2-51509 awarded by the Space Telescope Science Institute, which is operated by the Association of Universities for Research in Astronomy, Inc., for NASA, under contract NAS5-26555. E.R-R thanks the Heising-Simons Foundation and the NSF (AST-1911206, AST-1852393, and AST-1615881) for support.

This paper includes data collected with the TESS mission, obtained from the MAST data archive at the Space Telescope Science Institute (STScI). Funding for the TESS mission is provided by the NASA Explorer Program. STScI is operated by the Association of Universities for Research in Astronomy, Inc., under NASA contract NAS 5-26555.

\end{acknowledgements}

\noindent
{\it Facilities}: {\it Chandra}, {\it TESS}, ASAS-SN

\software{CIAO (v4.7; \citealt{fru06}), XSPEC (v12.10.1f; \citealt{arnaud96}), PHEOBE (v2.3, \citealt{conroy20})}
\nocite{*}
\bibliographystyle{aasjournal}
\bibliography{wr20a}

\end{document}